\documentclass[]{aa}  
\usepackage{graphicx}
\usepackage{txfonts}
\usepackage{natbib} 
\bibpunct{(}{)}{;}{a}{}{,} % to follow the A&A style

\begin{document}

\newcommand{\teff}{\ensuremath{T_{\rm{eff}}}}
\newcommand{\logg}{\ensuremath{\log g}}
\newcommand{\lheh}{\ensuremath{\log \left(N_{\rm{He}}/N_{\rm{H}}\right)}}
\newcommand{\nyvir}{\object{PG\,1336$-$018}}
\newcommand{\ra}{\ensuremath{\alpha_{\rm 2000}}}
\newcommand{\dec}{\ensuremath{\delta_{\rm 2000}}}
\newcommand{\vmag}{\ensuremath{m_{\rm V}}}
\newcommand{\msol}{\ensuremath{M_{\odot}}}
\newcommand{\rsol}{\ensuremath{R_{\odot}}}

\title{The binary properties of the pulsating\\ subdwarf B eclipsing binary
       \object{PG\,1336$-$018} (NY Vir)\thanks{
       Based on observations collected at the European Southern Observatory,
       Chile. Program ID: 075.D-0174.}
}

\author{M.~Vu\v{c}kovi\'{c}\inst{1} \and C.~Aerts \inst{1,2} \and
   R.~\O stensen \inst{1} \and G.~Nelemans\inst{2} \and H.~Hu \inst{1,2} \and
   C.S.~Jeffery \inst{3} \and V.S.~Dhillon \inst{4} \and T.R.~Marsh \inst{5} }

\offprints{maja.vuckovic@ster.kuleuven.be}

\institute{ Institute for Astronomy, K.U.~Leuven, Celestijnenlaan
          200D, 3001 Leuven, Belgium\\ \email{maja.vuckovic@ster.kuleuven.be}
    \and Department of Astrophysics, Institute of Mathematics, Astrophysics 
 and Particle Physics (IMAPP), Radboud University, 6500 GL Nijmegen, 
The Netherlands
          \and Armagh Observatory, College Hill, Armagh, Northern Ireland BT61
          9DG
	   \and Department of Physics and Astronomy, University of Sheffield,
          Sheffield S3 7RH, UK
          \and Department of Physics, University of Warwick, Coventry CV4 7AL,
          UK
	  }

\date{Received DD MM 2006 / Accepted 24 05 2007}

% \abstract{}{}{}{}{}
% 5 {} token are mandatory
\abstract{}{
%}{
  We present an unbiased orbit solution and mass determination of the components
  of the eclipsing binary \nyvir\ as a critical test for the formation scenarios
  of subdwarf B stars.  }{ We obtained high-resolution time series VLT/UVES
  spectra and high-speed multicolour VLT/ULTRACAM photometric observations of
  \nyvir, a rapidly pulsating subdwarf B star in a short period eclipsing
  binary.  }{ Combining the radial velocity curve obtained from the VLT/UVES
  spectra with the VLT/ULTRACAM multicolour lightcurves, we determined numerical
  orbital solutions for this eclipsing binary.  Due to the large number of free
  parameters and their strong correlations, no unique solution could be found,
  only families of solutions.  We present three solutions of equal statistical
  significance, two of which are compatible with the primary having gone through
  a core He-flash and a common-envelope phase described by the
  $\alpha$-formalism. These two models have an sdB primary of 0.466\,\msol\ and
  0.389\,\msol, respectively.  Finally, we report the detection of the
  Rossiter-McLaughlin effect for \nyvir.}{}

\titlerunning{The binary properties of the pulsating sdB eclipsing binary \nyvir}
\authorrunning{M.~Vu\v{c}kovi\'{c} et al.}

\maketitle

%
%________________________________________________________________

\section{Introduction}\label{sect:intro_PG1336}

The subdwarf B (sdB) stars are generally acknowledged to be core helium burning
stars with a canonical mass of approximately 0.5\,\msol. Their thin, inert
hydrogen envelope ($M_{\rm env}$\,$\lesssim$\,0.02\,\msol) places them on the
hot extension of the Horizontal Branch (HB), the so-called Extreme Horizontal
Branch (EHB).  Since the hydrogen envelope is too thin to sustain nuclear
burning, these stars will not go through the Asymptotic Giant Branch and
Planetary Nebula phases.  Instead, when their core helium has run out, they 
will enter a He-shell burning phase, where they expand and heat up, making them 
appear as sdO stars before they evolve directly onto the white dwarf cooling sequence.
Even though the models describing the future evolution of the sdB stars are 
generally accepted \citep[e.g.~those of][]{Dorman1993ApJ}, the current evolutionary 
state of the sdB stars is still poorly understood.  The fact that sdB stars must have 
lost almost all of their hydrogen layer at \emph{exactly} the same time when the
helium core has attained the minimum mass required for the helium flash to
occur, makes them enigmatic from an evolutionary point of view. To loose such an
amount of mass, they must suffer considerable mass loss during the red giant
branch (RGB) phase, and most probably also during the helium core flash.  The
most fundamental missing piece to our understanding of the evolution of the sdB
stars, apart from the physics during the helium core flash, is the nature and
physics behind this mass loss \citep{Fusi-Pecci1976A&A}.
   
In recent years it has been discovered that a significant fraction of sdBs are
in binaries.  \cite{Maxted_2001} found that about two-thirds of the sdB stars in
the field are members of binaries.  \cite{Napiwotzki2004SPY} found a binary
fraction of 40\% among stars in the SPY (Supernova type Ia Progenitor) survey
sample, while \cite{Morales-Rueda2006ECSurvey} found 48\% in a sample from the
Edinburgh-Cape (EC) survey. Many of the binary sdBs are found to be in short
period systems with periods from a few hours to several days, with companions
being either white dwarfs or M-dwarfs \citep{M-Rueda2003MNRAS}.  The peculiar
frequency of binarity has been an important constraint on evolutionary
population synthesis theory, and has led to the acknowledgment that the binarity
has to play a key role in the formation channels for sdB stars. There are
several binary mechanisms proposed by \citep[][and references therein]{Han2002,
Han2003} as formation channels for sdB stars : 
\begin{enumerate}
\item common envelope ejection, leading to short-period binaries with periods
      between 0.1 and 10 days and an sdB star with a very thin hydrogen
      envelope, and with a mass distribution that peaks sharply at 0.46\,\msol.
      Depending on the secondary, a main--sequence star or a white dwarf, the
      subchannels are called the first CE ejection channel and the second CE
      ejection channel, respectively,
      
\item stable Roche lobe overflow, resulting in similar masses as in 1.~but with
      a rather thick hydrogen-rich envelope and longer orbital periods between
      10 and 100 days,
      
\item double helium white dwarf mergers giving rise to single sdB stars with a
wider distribution of masses.
\end{enumerate}

Detailed investigation of sdB binaries is crucial in order to determine their
masses for comparison with the theoretically proposed evolutionary channels.
New momentum in the efforts to resolve the evolutionary paths of sdB stars came
a decade ago, after the discovery that some of them pulsate
\citep{Kilkenny1997}. This has opened up a new window into their interiors via
the techniques of asteroseismology and stimulated a burst of research.
Extensive search campaigns have revealed two classes of pulsating sdB stars
known as short period sdB variables (sdBV or V361\,Hya stars, formerly EC\,14026
stars, after the prototype) and long period sdB variables known as PG\,1716
stars \citep[or lpsdBV stars,][]{Green2003}.
   
The sdBV stars, discovered by \cite{Kilkenny1997} and independently
theoretically predicted by \cite{Charpinet1996}, are low amplitude multimode
pulsators with typical periods ranging between 100--250\,s.  Their pulsation
amplitudes are generally of the order of a few hundredths of a magnitude. The
short periods, being of the order of and shorter than the radial fundamental
mode for these stars, suggest that the observed modes are low-order, low-degree
$p$-modes \citep{Charpinet2000}. The 39 known sdBV stars occupy a region in the
\teff\,--\,\logg\ plane with effective temperatures between 28\,000\,K and
36\,000\,K and surface gravities (\logg) between 5.2 and 6.2.
   
The detailed asteroseismological modelling of sdBV stars is hampered by the fact
that there are too few pulsational frequencies to fit those predicted from
non-rotating or rigidly rotating models
\citep{Brassard2001ApJ,Charpinet2005A&A,Randall2005ApJS}. However, the observed
frequency spectra are too dense to be accounted for by only low-degree
($\ell$\,$\leq$\,2) modes. In order to have a unique asteroseismological model
we need to have accurate pulsation frequencies \emph{and} an unambiguous
identification of the modes of oscillation (spherical wavenumbers $\ell$ and
$m$). Thanks to multisite campaigns by the
WET\footnote{http://wet.physics.iastate.edu/} devoted to resolving the frequency
spectrum of sdBV stars in the last decade, we do have extensive and reliable
frequency lists for several sdBVs.  The problem lies in the second requirement
mentioned above, the unambiguous mode identification.  There are only two ways
this can be achieved: through line profile variations \citep{Aerts2000} or the
amplitude ratio method \citep{Dupret2003,Randall2005ApJS}.
   
%%%%% Two column figure (place early!)
\begin{figure*}
\centering
\rotatebox{-90}{\resizebox{8cm}{!}{\includegraphics[bb=50 50 370 760]{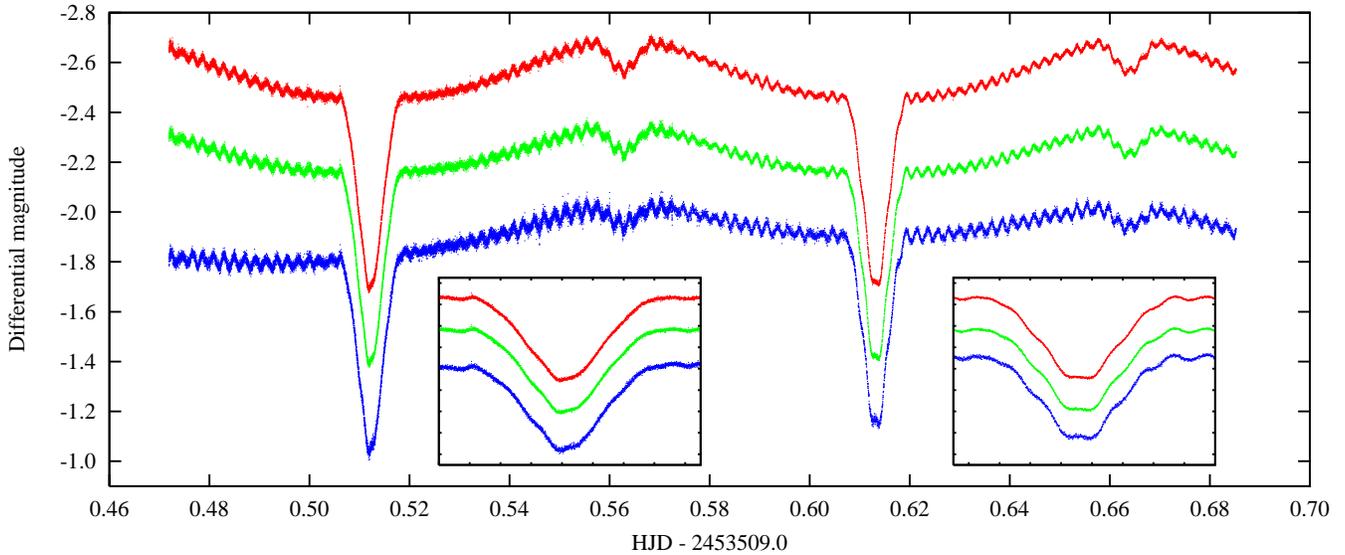}}}
\caption{ ULTRACAM/VLT $r$' (upper), $g$' (middle) and $u$' (bottom) lightcurves
of the eclipsing sdBV star \nyvir \, from 2005 May 18/19.  The insets show
enlarged sections of the two primary eclipses, where pulsations are clearly
visible. The differences between the two consecutive primary eclipses, apart
from the noise, are due to the beating of the modes and different phases covered
during the eclipse. The shape of the $u$' lightcurve is discussed in the text.
The ordinate is the differential magnitude, and the abscissa is Fractional
Julian Date.}
\label{Fig_gLC}
\end{figure*}

As sdBV stars are quite faint (the brightest one is $m_{\rm B}$=11.8) and their
periods are very short, the line profile variation method poses a real challenge
considering the low S/N that accompanies any high-resolution time-resolved
spectroscopy, even with the biggest telescopes available.  Hence, the line
profile variation method has not yet been reliably applied to any sdBV star.
The amplitude ratio method is not problem free either.  Due to the very low
pulsational amplitudes, the photometric errors are usually too large for
unambiguous identification of the spherical degree $\ell$ of the modes,
especially to distinguish between $\ell$= 0, 1 and 2 modes \citep{Jeffery2005}.
 
Among the binary sdB stars, four eclipsing sdB systems have been discovered that
all show a deep and strong reflection effect, with very short orbital periods in
the rather narrow range of 130--170\,minutes. Such short orbital periods imply that
they must have evolved through binary mass transfer and common envelope
evolution.  Out of these four systems, namely \object{HW Vir} \citep{Wood1993,
Menzies1986}, NY Vir \citep{D.Kilkenny1998MNRAS} (hereafter \nyvir),
\object{HS\,0705+6700} \citep{Drechsel2001A&A} and \object{HS\,2231+2441}
\citep{Roy2007}, only \emph{one} system contains a rapidly pulsating sdB star as
a primary: \nyvir. As such, this system provides a natural laboratory for
detailed evolutionary and asteroseismic analyses, which is the purpose of our
project.
    
\nyvir\ was classified as an sdB star in the Palomar--Green survey
\citep{GreenSchmidtandLiebert86} and shown to be a close eclipsing binary with
short-period multimode light variations by \cite{D.Kilkenny1998MNRAS}.
\emph{Assuming} the primary mass to be the canonical sdB mass of 0.5\,\msol,
\cite{D.Kilkenny1998MNRAS} find that the secondary must be a mid--M dwarf with a
mass of about 0.15\,\msol. Soon after its discovery, \nyvir\ was a target of two
Whole Earth Telescope (WET) campaigns, Xcov 17 in April 1999
\citep{D.Kilkenny2003MNRAS} and Xcov 21 in April 2001. These white light data
resolved more than 20 frequencies in the temporal spectrum
\citep{D.Kilkenny2003MNRAS} in the range from 5000 to 8000\,$\mu$Hz. Even though
the frequency content of the star is thus known very precisely, an adequate
asteroseismic model is still lacking mainly due to the lack of an unambiguous
mode identification.  The colour behaviour is needed for photometric mode
identification to identify the spherical degree $\ell$ of the modes and to
discriminate between the numerous possible seismic models. To further reduce the
allowable seismic model space we need to examine line profile variations due to
the pulsations in order to disentangle the azimuthal wavenumber $m$. Only with
the accurate pulsation frequencies and an \emph{unambiguous} mode identification
can the asteroseismology provide the accurate mass estimate needed for
confrontation with those predicted from the formation scenarios for sdB stars.
    
\nyvir, being the \emph{only} rapidly pulsating sdB star in an eclipsing binary,
is the only star with enough potential to confront the proposed evolutionary
scenarios, as the eclipses help constrain the inclination and radii.
Therefore we study \nyvir, this time armed with new multicolour photometric and
spectroscopic VLT data. In this paper we present the new data and the orbital
solution.  This is the first step toward our ultimate goal, an accurate mass
determination of \nyvir\ and a critical test of current stellar evolution
theory.

%__________________________________________________________________

\section{Observations and data reduction}\label{sect:observations_PG1336}

\subsection{Photometry}

\nyvir\ (\ra\,=\,13:38:48.2, \dec\,=\,--02:01:49.0, \vmag\,= 13.4) was observed
on the night of May 18/19 2005 using the ULTRACAM camera attached to the ESO VLT
UT3 (Melipal) at Paranal Observatory, Chile. ULTRACAM is a high-speed
three-channel CCD camera specifically designed for fast photometry programmes
\citep{Dhillon&Marsh2001}. We gathered two full orbital cycles, about 5\,h, of
\nyvir\ simultaneously in three filters $r$', $g$' and $u$' of the SDSS system
\citep{Fukugita1996}.  The seeing (around 0.9 arcsec) was variable during the
night and getting worse toward the end of the run. The exposure time was 0.5\,s
in the beginning of the run, but due to poorer seeing was increased to 1\,s to
improve the S/N. This did not deteriorate our temporal resolution significantly,
since the shortest period found in \nyvir\ is 97\,s
\citep{D.Kilkenny2003MNRAS}. To achieve 1 second time resolution, it was
necessary to define 2 windows on each of the 3 ULTRACAM chips. One window was
placed around PG1338-018, and another on a nearby comparison star. The
dead--time of the observation was 24 milliseconds.

All data frames were reduced using the ULTRACAM pipeline reduction software
\citep{Dhillon&Marsh2001}.  Care was taken to select the most optimal choices
offered in the reduction software. The 'normal' extraction method with the
'variable' aperture sizes, as they track local changes in the seeing disk, gave
the best results. Several apertures were tried out and an aperture of 1.7 times
the FWHM gave the highest S/N for $r$' and $g$' band. The star counts were
divided by the comparison star counts and converted to obtain a differential
magnitude (V--C) in each filter. As both the target and the comparison star were
in the same field, differential photometry accounted well for the variations in
the sky transparency and extinction in $r$' and $g$' band. Unfortunately, the
only comparison star within ULTRACAM's 2.6 arcminute field of view on the VLT is
very faint in the blue, resulting in poorer differential photometry in the $u$'
compared to the $r$' and $g$' band. Therefore, a wider aperture had to be used
for the $u$' band. Due to the faintness of the comparison star in $u$', its $g$'
band lightcurve was used to make the differential $u$' lightcurve. This gave a
satisfactory result in the sense that both the pulsations and the eclipses were
recovered, but it introduced an unreliable slope in the first part of the $u$'
lightcurve (see Fig.~\ref{Fig_gLC}). Therefore, we did not rely on the $u$'
lightcurve for the orbital analysis. However, we did use the second part of the
$u$' lightcurve to cross--check our results, as well as for the frequency
analysis (see Sect.~\ref{subsect:binarity and pulsation}).

The times in the data frames were converted to JD and barycentrically
corrected. Differential (V--C) lightcurves for $r$', $g$' and $u$' were
constructed from a set of more than 80\,000 science frames.  The $r$' , $g$' and
$u$' lightcurves are plotted in Fig.~\ref{Fig_gLC}, where we can see a clear
sign of the pulsations of the primary component in \emph{all} the phases of the
binary orbit, even during the primary eclipse.  A strong reflection-like effect
(0.2 magnitudes in $g$' and 0.25 magnitudes in $r$') is evident.  This effect,
characteristic of all binary systems containing an sdB star and a cool M--dwarf
companion in rotationally locked orbit, is due to the high contrast in the
temperatures between the heated and unheated hemispheres of the M--dwarf.

%______________________________________________ 
  
\begin{figure}
\centering
\includegraphics[angle=-90, width=\hsize]{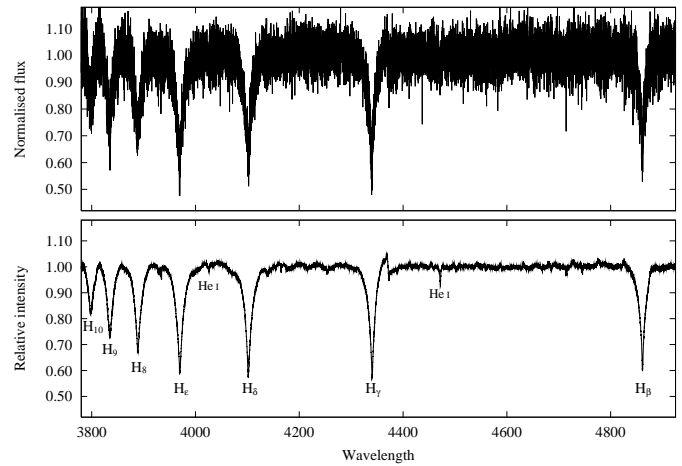}
\caption{A typical single UVES/VLT spectrum of \nyvir\ from our VLT run on 2005
April 28 (top) and the coadded spectrum (bottom), produced by combining all the
399 available spectra after shifting according to the orbital radial velocity
solution.  The Balmer lines are indicated together with the helium lines used
for the determination of physical parameters. Discontinuities due to imperfect
merging of spectral orders only become evident in the high--S/N combined
spectrum.}
\label{Fig_spectrum}
\end{figure}

%______________________________________________ 

\subsection{Spectroscopy}
 
Even though \nyvir\ was a target of several photometric campaigns, its faintness
relative to the rapid oscillations has prevented any reasonably good
time-resolved spectroscopy. The short pulsation periods require very short
integration times. There were two attempt so far with the aim of detecting the
pulsational radial velocities \citep{Woolf2003} and identification of the
pulsation modes from the wavelength dependency of the amplitudes
\citep{Dreizler2000BaltA}, both with a null result.

 A time-series of 399 high resolution spectra were taken over a period of
 $\sim$\,9\,h, covering about 3.7 full orbits, on the night of April 28, 2005
 using the Ultraviolet Visual Echelle Spectrograph (UVES) on the VLT UT2
 (Kueyen) at the Paranal Observatory, Chile.  Only the blue arm was used, with
 wavelength coverage from 3900 to 5000\,\AA, and the slit width of 1 arcsec at a
 resolution of 46\,890.  Each spectrum was integrated for 45\,s which, with the
 ultra fast read-out of about 23\,s we used, gave a time resolution of 68\,s.
 Dome flat-fields and bias calibration frames were taken at the beginning and at
 the end of the night, and ThAr exposures were taken before and after the run.

Due to the very low signal we got for such a short exposure and the ultra fast
read-out mode used, the UVES reduction pipeline did not give satisfactory
results. Therefore, we developed a non-standard reduction method, using the
ESO-MIDAS package. This provided a factor of $\sim$\,2 increase in the S/N ratio
of the reduced spectra, compared to those produced by the pipeline.  The bias
calibration frames had an offset between the upper and the lower part, due to
the ultra fast read-out mode used.  After careful examination of each bias
frame, we proceeded as follows.  First we examined the interorder space of each
science frame (by taking the median of the box) to determine these offsets which
were then subtracted from the science frames. Then the science frame was
corrected for cosmic rays, extracted and background corrected (which was
smoothed to reduce the noise). Since, in our case, the sky background
contributes most to the noise, we used optimal extraction which gave better S/N,
as suggested by \cite{Mukai1990}. Then the science frames were flat-field
corrected, wavelength calibrated and, finally, the orders were merged. Since the
spectra were oversampled we have rebinned them in an optimal way such that the
S/N increased without compromising the resolution. Finally, the science frames
were normalized.

%______________________________________________ 
  
\begin{figure}[t]
\centering
\includegraphics[angle=-90, width=\hsize]{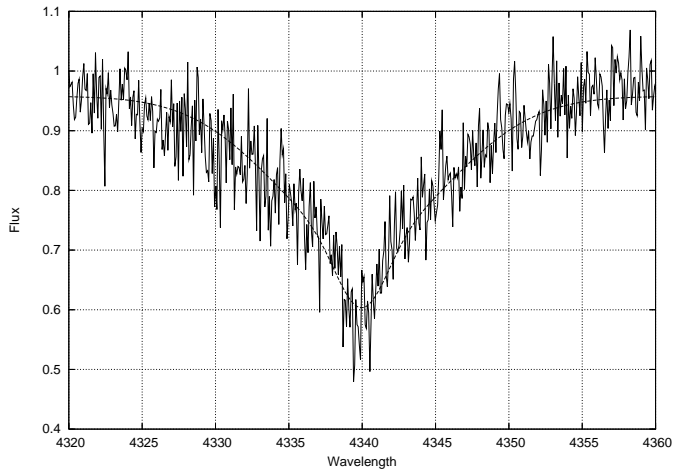}
\caption{A sample fitting of two Gaussians to the observed H$_\gamma$ line (the
same spectrum as the one shown in Fig.~\ref{Fig_spectrum}) using {\tt molly}.  }
\label{Fig_mollyfit}
\end{figure}

%______________________________________________ 

A typical individual spectrum of \nyvir\ is shown in the top panel of
Fig.~\ref{Fig_spectrum}.  The bottom panel of Fig.~\ref{Fig_spectrum} shows the
coadded orbit-corrected spectrum (see Sect.~\ref{sect:specfit}). Despite our
extensive effort to achieve the optimal reduction scheme, the extraction and
merging of the orders is not perfect.  This is due to the fact that the Echelle
order discontinuities do not behave 'consistently' under a low signal.  This
leads to some jumps and wiggles seen in the continuum of the coadded spectrum
and particularly in the red wing of H$_\gamma$. For this reason we did not make
use of this line in the merged spectrum for the spectroscopic parameter
determination discussed below.

 In the blue wavelength range covered by our data no sign of any spectral
feature from the cool companion can be seen, confirming the results of
\cite{Woolf2003}. Due to the large difference in effective temperatures (about a
factor of 10, see Sect.~\ref{sect:orbit_PG1336}) the hot sdBV star dominates the
spectrum even in the primary eclipse.
      
%__________________________________________________________________

\section{RV determination}\label{sect:RV_PG1336}

Our spectra allow us to produce a radial velocity (RV) curve, with an excellent
phase coverage, from which we can independently determine the orbital period
($P$) and semi-amplitude ($K_1$) of this eclipsing binary.  As we are dealing
with a low S/N, we determined RVs from the spectra trying out several different
methods. The best results were obtained by using {\tt molly}- a software
package, which fits two Gaussian profiles to the Balmer line profiles
\footnote{http://deneb.astro.warwick.ac.uk/phsaap/software/molly/html/IN\-DEX.html}.
This allows good treatment of both the broad wings and the sharper core at the
same time.  This gave better results than any of the other methods we have
tried.
 
We have measured the RVs of the highest S/N lines in the spectrum, namely
H$_{\epsilon}$, H$_{\delta}$, H$_{\gamma}$ and H$_{\beta}$, using this package.
A sample fit is shown in Fig.~\ref{Fig_mollyfit} for an individual spectrum. The
FWHM of the two Gaussian fits, as well as their heights, were treated as a free
parameter at first, but were kept fixed once the best fit values were found. We
checked carefully if the RV from the H$_{\gamma}$ line deviated from the one of
the other Balmer lines, due to the discontinuity in its red wing. This turned
out not to be the case (see also Fig.~\ref{Fig_mollyfit}) so we kept the
H$_{\gamma}$ RV values in our analysis.

\begin{figure}
\centering \includegraphics[angle=-90, width=\hsize]{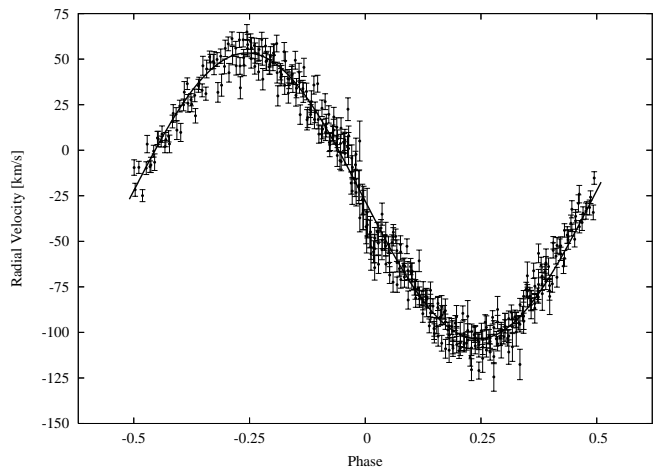}
\caption{The radial velocity measurements (average of the H$_{\rm \epsilon}$,
H$_{\rm \delta}$, H$_{\rm \gamma}$ and H$_{\rm \beta}$ lines) of all the
individual UVES/VLT spectra. The best fit orbit solution from {\tt PHOEBE} is
also shown.}
\label{Fig_rv}
\end{figure}

Finally, the average of each RV measurement, using H$_{\epsilon}$,
H$_{\delta}$, H$_{\gamma}$ and H$_{\beta}$ lines, was determined. These radial
velocity values for each of the 399 individual spectra (with the errors), are
shown in Fig.~\ref{Fig_rv} together with the best fit orbital solution (see
Sect.~\ref{sect:orbit_PG1336}). 

%______________________________________________ 
  
%__________________________________________________________________

To perform an independent determination of the orbit from our spectroscopic
data, and to verify the photometric ephemeris, the measured RVs (after
barycentric correction of the velocities and the mid-exposure times) were
subjected to a periodogram analysis. A sinusoidal fit using Period04
\citep{Period04} gives the frequency 114.25\,$\pm$\,0.1\,$\mu$Hz and the
semi--amplitude 78.6\,$\pm$\,0.6\,km/s which is, considering our poor frequency
resolution of about 30\,$\mu$Hz, in a good agreement with the orbital period P=
0.101015999 \,d calculated by \cite{D.Kilkenny2000Obs} as well as with the
values derived from {\tt PHOEBE} in Section \ref{sect:orbit_PG1336}. The
semi--amplitude of the velocity variation is in good agreement with the
78\,$\pm$\,3\,km/s estimated by \cite{D.Kilkenny1998MNRAS} (see their Table 4)
even though they reported the semi--amplitudes of all of their observations (see
their Table 3) to range from 47\,$\pm$\,4\, to 79\,$\pm$\,4\,km/s. The
semi--amplitude we obtained is somewhat larger than estimated by
\cite{Woolf2003}, 64\,$\pm$\,1\,km/s, but their data cover only 1.4 orbits and
contain a gap which probably resulted in an underestimated value.

As our data set suffers from a baseline too short for reliable ephemeris
determination, we adopted the ephemeris obtained by \cite{D.Kilkenny2000Obs}
(see Table \ref{tbl:phoebe_fixed_parameters}).

Since the system is single-lined and the orbit is assumed to be circular, the
analysis of the RV curve is straightforward. The mass function calculated from
the semi-amplitude and the period gives:
\begin{eqnarray*}
     f(M) = 0.0051 \pm 0.0001\,M_{\odot}\,.
\end{eqnarray*}

%__________________________________________________________________

\section{Orbital parameters}\label{sect:orbit_PG1336}

In order to investigate the pulsational properties of \nyvir, the subject of a
follow-up paper, the orbital variations due to the binarity must be removed from
the observed lightcurve. However, in order to find the best orbital solution for
this eclipsing binary system, the pulsations of the sdB primary must be removed
as well. This is a non-trivial coupled problem.  The determination of the
orbital parameters of this system required to understand and evaluate the
temporal spectrum of the primary sdB pulsator. In order to achieve this, we
followed an iterative procedure, using all the information about the target we
have. Once we find a reliable orbital solution, we subtract it from the
lightcurves. Then we use the orbit subtracted lightcurves to extract the
pulsation frequencies present in our data. We prewhiten the original observed
lightcurves with these frequencies.  The prewhitened lightcurves are then used
as input to find the second iteration orbital solution.

%______________________________________________ 
  
\begin{figure}
\centering
\includegraphics[width=7cm]{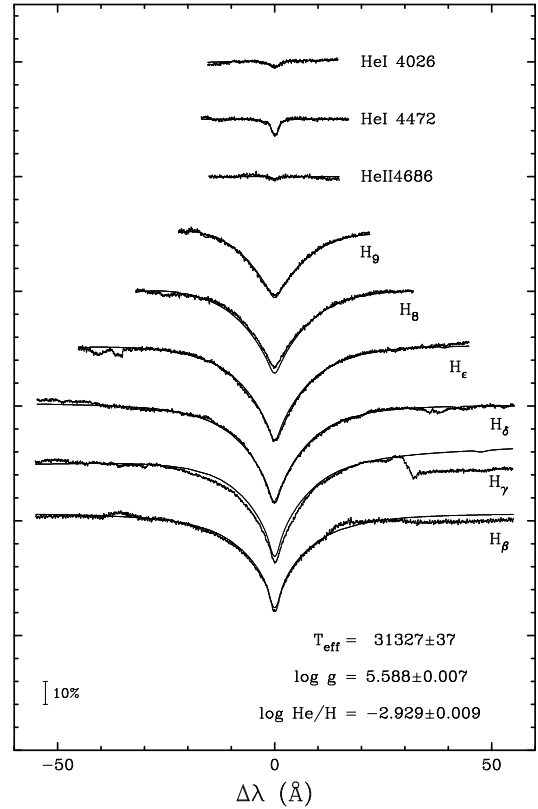}
\caption{
Our spectroscopic model fit to the mean spectrum in Fig.~\ref{Fig_spectrum}.
The best fit model spectrum has been plotted on top of the observed spectrum
as a smooth curve. Note that the \ion{H}{$\gamma$} line was kept out of the
fit due to its proximity to an echelle order discontinuity.
}
\label{Fig_spfit}
\end{figure}

%__________________________________________________________________

\subsection{Fundamental parameters}\label{sect:specfit}

Our high resolution VLT/UVES spectra allow us to improve the spectroscopic
parameters determined by \cite{D.Kilkenny1998MNRAS}.  Using our RV solution (see
Fig.~\ref{Fig_rv}), we shifted the spectra and added them together to improve
the S/N. The coadded orbit-subtracted spectrum is shown in the bottom panel of
Fig.~\ref{Fig_spectrum}.

For the model fitting procedure, we used the LTE models of \cite{Heber2000}.
The model spectra were convolved with a Gaussian instrumental profile of
0.25\,\AA\, and rotationally broadened (assuming tidally locked rotation) with a
$v\sin i$ of 74.2\,km/s. This produces a model spectrum with line cores that
reproduce the observed spectrum excellently for all lines that are unaffected by
Echelle order discontinuities.  Unfortunately, while the fit to the cores is
good, the wings are not well fitted. Our best simultaneous fit for effective
temperature, gravity and helium abundance yields:
\begin{eqnarray*}
T_{\rm eff} & = & 31300 \pm 250\,{\rm K}\\
\log g      & = & 5.60 \pm 0.05\,{\rm dex}\\
\log y      & = & -2.93 \pm 0.05\,{\rm dex}
\label{loggteff}
\end{eqnarray*}
The quoted errors are about five times larger than the formal fitting
errors reported in Fig.~\ref{Fig_spfit}.
Although such 5$\sigma$ errors would normally be quite conservative considering
the resolution and signal of the combined spectrum, there are obvious
problems. The effects of errors due to the Echelle extraction
problems described earlier are hard to quantify.  The effective temperature is
well constrained by the depth of the high order Balmer lines, and the helium
abundance is determined by the depth of the narrow \ion{He}{i} lines (marked in
Figs.~\ref{Fig_spectrum} and \ref{Fig_spfit}), 
which are not much affected by the Echelle extraction problems. However, since
the Echelle order discontinuities strongly affect the wings of the lines, which
are essential for the gravity determination, we cannot exclude a large error on
\logg.  For this reason, we will only use the effective temperature
determination as a constraint for our orbital fitting procedure, and not \logg.
Indeed, as we will see later, such a low \logg\ is inconsistent with any
realistic mass--radius relationship that can be derived from the orbit by at
least 0.15 dex. In order to rule out other causes for the inconsistent \logg
determination from the average spectrum, we tried
to fit it using NLTE atmosphere models, enhanced metallicity models, or
changing the assumed rotational velocity broadening. All these attempts produced
negligible changes to the derived parameters listed above. 

%______________________________________________ 

\begin{figure*}
\centering
\includegraphics[angle=-90, width=18cm]{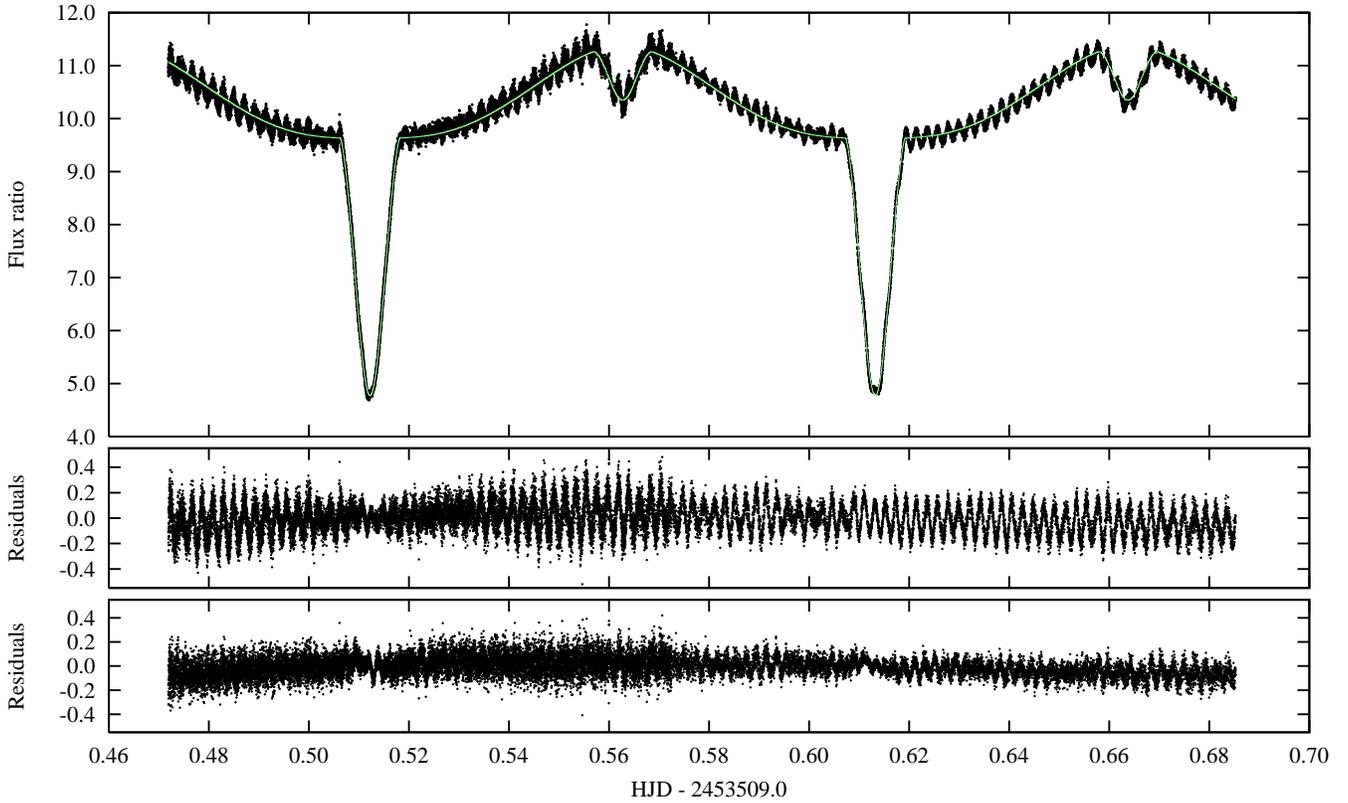}
\caption{The ULTRACAM/VLT $g$' lightcurve together with the synthetic orbit
solution.  The middle panel shows the residuals of the orbit
subtraction. Pulsations during the eclipses are now clearly visible, and we can
see that the amplitude is smaller during the primary eclipse than during the
secondary as only the part of the surface is visible. The bottom panel shows the
residuals after prewhitening with the four strongest oscillation modes. }
\label{Fig:Gfit}
\end{figure*}

%__________________________________________________________________

\subsection{Binarity and pulsation}\label{subsect:binarity and pulsation}

Numerical orbit solutions were investigated using the {\tt PHOEBE} package tool
\citep{Prsa2005} which incorporates the 
aspects of the Wilson--Devinney (WD) code \citep{WilsonandDevinney1971}.  The WD
approach uses differential correction (DC) as the minimization method, which is
in essence a linearised least squares method.  The code was used in the mode for
detached binaries with no constraints on the stellar potentials.  No third light
or spots were included.

The ULTRACAM/VLT $g$' and $r$' lightcurves and the RV measurements obtained from
the UVES/VLT spectra were solved simultaneously to yield a consistent model
fit. As {\tt PHOEBE} is limited by the number of points (currently the limit is
9000 points) we had to phase bin our ULTRACAM/VLT lightcurves into 4000 data
points per lightcurve.

The major problem in finding the orbital solution of any binary system is not
only the fact that there are many free parameters (12 + 5$n$, where $n$ is the
number of lightcurves in different filters), but also that the parameters are
correlated. Some of these correlations are severe, especially between the mass
ratio $q$ and the potential of the secondary star $\Omega_2$ (see the discussion below in
Sect.~\ref{sect:discussion_PG1336}). Hence, one is left with several formal
families of solutions within the parameter space. We must then confine the range
of possible solutions by reducing the number of free parameters. The only safe
way to do this is by considering the boundary conditions set by the data
themselves and by sound theoretical considerations.
 
The parameters that were assumed and kept fixed in our analysis were $t_0$, P,
$T_{\rm eff}$ of the primary, gravity darkening coefficients both for the
primary $g_1$ and the secondary $g_2$, bolometric albedo of the primary $A_1$ and the limb
darkening coefficients of the primary in the two filters $x_1$ ($g$', $r$'). For the gravity
darkening coefficients we adopted values of 1.0 for the primary (radiative
envelope) and 0.32 for the secondary (convective envelope). We assumed a
circular orbit ($e$=0) and synchronized rotation with the orbit.

%__________________________________________________ One column table

\begin{table}
\caption[]{Fixed parameters in the search for the orbital solution of \nyvir.}
\label{tbl:phoebe_fixed_parameters}
\centering
\begin{tabular}{ll} \hline
Parameter       &  Value  \\ \hline
$t_0$           & 2450223.36134\,d$^{\mathrm{a}}$     \\
$P$             & 0.101015999\,d$^{\mathrm{a}}$       \\
$T_{\rm eff1}$  & 31300\,K                       \\
$T_{\rm eff2}$  & 3000\,K$^{\mathrm{b}}$         \\
$g_1$  &     1.0   \\
$g_2$  &    0.32   \\
$A_1$  &     1.0   \\
$x_1$ ($g$') &    0.217  \\
$x_1$ ($r$') &    0.178  \\ \hline
\end{tabular}
\begin{list}{}{}
\item[$^{\mathrm{a}}$] Ephemeris taken from \cite{D.Kilkenny2000Obs}.
\item[$^{\mathrm{b}}$] $T_{\rm eff2}$ was kept fixed as it is poorly constrained
by the data, see the text for details.
\end{list}
\end{table}

\begin{figure*}
\centering
\includegraphics[angle=-90, width=18cm]{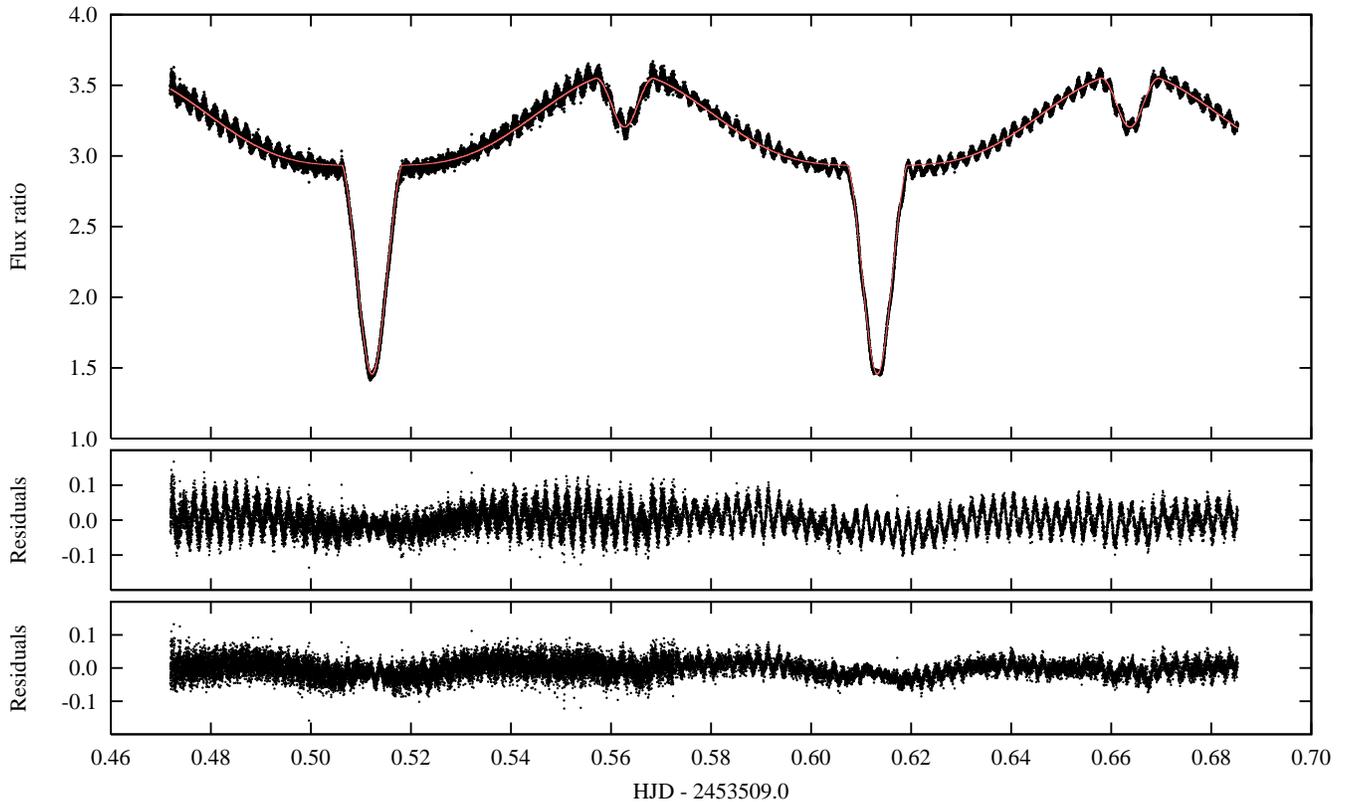}
\caption{Same as Fig.~\ref{Fig:Gfit} but for the $r$' band.
The trends seen in the middle and bottom panel result from imperfect removal
of the reflection effect due to the changing temperature across the surface
of the secondary (see text for details). }
\label{Fig:Rfit}
\end{figure*}

The effective temperature of the primary $T_{\rm eff1}$ was set to the value
derived from our spectra (see Sect.~\ref{sect:specfit}). 
The effective temperature previously estimated by \citet[][ $T_{\rm eff} = 33\,000 \pm 1\,000$]{D.Kilkenny1998MNRAS} 
was used as well, but, as it did not influence the derived parameters except 
for the luminosity of the stars, we fixed the temperature to the value derived 
by our new data.  The $T_{\rm eff2}$ of
the secondary has a very low contribution to the total flux (see
Sect.~\ref{sect:observations_PG1336}) and, therefore, is not tightly
constrained. An appropriate treatment of the effective temperature of the
secondary in the case where the hot sdB primary is heating the cool secondary is
not trivial, as the temperature on the illuminated hemisphere can be as much as
five times higher than on the non-illuminated one \citep{S.Zola2000BaltA}.
Whilst we did not intend to fix the effective temperature of the secondary star
at first, we have found that leaving it as an adjustable parameter does not give
consistent results.  With $T_{\rm eff2}$ as a free parameter, it converges to
around 4000\,K for the $g$' lightcurve, but to only 2700\,K for the $r$'-band
lightcurve.  As a reasonable compromise for $T_{\rm eff2}$, we choose to fix it
to 3000\,K.  Considering the fact that the contribution of the secondary to the
total flux is negligible, this is not an obstacle. 

As there are no published limb darkening coefficients for sdB stars we 
calculated the limb darkening coefficients $x_1$ ($g$', $r$' and $u$') for a
'typical' sdB star from a fully line--blanketed LTE model atmosphere
\citep{Behara2006} with $T_{\rm eff}$ = 30\,000\,K, $\log g$ = 5.5, $V_{\rm
turb}$=5\,km/s and solar abundances (a linear cosine law was used). The mean
limb darkening coefficients in each filter were computed by convolving the {\tt
ULTRACAM} efficiencies in each filter with the monochromatic limb darkening
coefficients and the stellar fluxes. We also computed the orbital solution using
an extrapolation of previously reported coefficients from the tables of \cite{Wade1985} and 
\cite{Al-Naimiy1978}, as well as the values fixed at 0.25 (V) and 
0.20 (R) \citep{D.Kilkenny1998MNRAS}. This did not change the solution, so we adopted the
coefficients we computed from a modern atmosphere model.
Table~\ref{tbl:phoebe_fixed_parameters} summarises the values of the fixed
parameters.  The surface gravity is not a free parameter obtained by {\tt
PHOEBE}, since it is defined by the mass and radius.

Using the ephemeris given in \cite{D.Kilkenny2000Obs} we find a phase shift
of 0.00374$\pm$0.00006 d. This phase shift could in principle be due to timing
errors in our data rather than to an intrinsic change in the system. However, we
carefully checked timings in our data sets and, moreover, we have data from two
different instruments which both show the same phase shift. A timing error is
therefore very unlikely to be the cause of the measured shift.  A change
inherent to the system is thus the most probable reason.  With only two
minima timings we cannot draw any further conclusion here, only emphasise the
need for further epoch observations.  A similar period change on the order of
0.003\,d over a period of 6 years in the \object{HW Vir} system was documented
by \citet{D.Kilkenny2000Obs}.  

The strong pulsations in the lightcurves are obstructing the fine tuning of the
orbit, as the pulsations are seen as scatter by {\tt PHOEBE}. Therefore, we take
the first iteration solution and subtract it from the lightcurves. Now, after
the dominant parts of the periodicity, i.e. the eclipses, have been removed from
the lightcurves we can analyse them in order to take out the pulsations of the
primary from the lightcurves.

%__________________________________________________________________

\begin{table*}
\caption{The list of frequencies, periods, amplitudes and phases we detected and
prewhitened our data with.  The phase is given as the time of maximum amplitude
since $t_0$.  }
\label{tbl:frequencies}
\centering \small\begin{tabular}{crrrrrrr}\hline\hline
\multicolumn{1}{c}{Frequency} &
\multicolumn{1}{c}{Period} &
\multicolumn{3}{c}{Amplitude}  &
\multicolumn{3}{c}{Phase ($T_{\rm max}$)} \\ 
\multicolumn{1}{c}{[$\mu$Hz]} &
\multicolumn{1}{c}{[s]} &
\multicolumn{3}{c}{[mma]}  &
\multicolumn{3}{c}{[s]} \\ \hline
\multicolumn{1}{c}{} &
\multicolumn{1}{c}{} &
\multicolumn{1}{c}{$g$'}  &
\multicolumn{1}{c}{$r$'}  &
\multicolumn{1}{c}{$u$'}  &
\multicolumn{1}{c}{$g$'}  &
\multicolumn{1}{c}{$r$'}  &
\multicolumn{1}{c}{$u$'}  \\ 
\hline
  5430.1 & 184.16 & 11.2(1) & 10.5(1) & 17.1(2) & 142.3(3)  & 142.2(3) & 
141.4(4) \\
  5579.9 & 179.21 &  3.8(1) &  3.7(1) & 3.5(2) & 105.9(8)  & 105.8(8) & 
115(2) \\
  5757.3 & 173.69 &  1.7(1) &  1.7(1) & 2.8(2) & 148(2) & 147(2) & 155(2) \\
  7076.7 & 141.31 &  2.0(1) &  1.9(1) & 3.0(2) & 105(1) & 106(1) & 107(2) \\ 
\hline
\end{tabular}
\end{table*}

%__________________________________________________________________

A Fourier amplitude spectrum was calculated for each orbit subtracted lightcurve
to deduce the periodicities present in the data. The short timespan of our
photometric data confines us with a frequency resolution of 54\,$\mu$Hz.  Since
we are unable to resolve many of the closely spaced frequencies in the spectrum
published by \cite{D.Kilkenny2003MNRAS}, we cannot use their peaks.  We can only
remove the periodicities we observe in our data in order to improve our orbit
solution, after verifying that the frequencies we detect are indeed in the range
of known \nyvir\ frequencies.
  
After identifying the highest amplitude peak in the spectrum and cross-checking
if this frequency is present in the previous data sets within our frequency
resolution, we remove this peak from the data by subtracting a sine wave (with
the frequency, amplitude and phase determined by a non-linear least-squares fit
-NLLS) from the original lightcurves. We calculate the Fourier amplitude
spectrum of the prewhitened residuals and repeat the procedure until no new
peaks could be securely identified.  In this way we are able to remove four
frequencies, as listed in Table~\ref{tbl:frequencies}.  The frequency spectrum
of \nyvir\ is complicated as there are many frequencies in a narrow frequency
range, which are unresolved in our data set. Therefore the NLLS would not
converge on a simultaneous fit to more than four frequencies, even though there
is still significant power left in the Fourier spectrum. That is also the reason
why the amplitudes appear higher in our data set compared to the ones seen in
\cite{D.Kilkenny2003MNRAS} as several frequencies are blended into one. The
highest amplitude frequency in our data set at 5430.1\, $\mu$Hz is most probably
the result of seven unresolved closely spaced frequencies $f_3, f_4, f_{25},
f_{10}, f_5, f_7$ and $f_{22}$ from Table 4 of \cite{D.Kilkenny2003MNRAS}.

These prewhitened lightcurves were then phase binned and, together with the RV
curve, fed into {\tt PHOEBE} to search for the improved orbit solution.  Even
though residual pulsations are still clearly visible in the lightcurves, their
amplitudes are now significantly smaller, which allows us to obtain a more
reliable (second iteration) orbit solution.  A third iteration step turns out to
be unnecessary, as it does not improve the final outcome of the orbital
parameters.

\begin{table}
\caption{System parameters of the three best model fits to RV data and
lightcurves of \nyvir.  The formal 1$\sigma$ error on the last digit of each
parameter is given in parentheses.  }
\label{tbl:system_param}      
\centering          
\begin{tabular}{rccc}     % 4 columns 
\hline\hline       
Free parameter & Model I & Model II & Model III \\
\hline                    
$a$ [\rsol]     &    0.723(5)   &    0.764(5)   &    0.795(5) \\
$q$             &    0.282(2)  &    0.262(2)  &    0.250(2)  \\
$i$ [$^\circ$]  &   80.67(8)   &   80.67(8)    &   80.67(8)    \\

$\Omega_1$      &    5.50(3)   &    5.48(3)   &    5.47(3)   \\
$\Omega_2$      &    2.77(1)   &    2.68(1)   &    2.62(1)   \\

      $A_2$ &    0.92(3)   &    0.92(3)   &    0.93(3)   \\
$x_2$ ($g$')& 0.38(8)     &    0.39(8)    &    0.38(8)    \\
$x_2$ ($r$')&  0.88(8)    &    0.89(8)    &    0.89(8)    \\
\hline  
\multicolumn{4}{l}{Derived parameters: } \\
\hline  

$M_1$ [M$_\odot$] & 0.389(5) & 0.466(6) & 0.530(7) \\
$M_2$ [M$_\odot$] & 0.110(1) & 0.122(1) & 0.133(2) \\ 
$R_1$ [R$_\odot$] & 0.14(1)  & 0.15(1)  & 0.15(1)  \\
$R_2$ [R$_\odot$] & 0.15(1)  & 0.16(1)  & 0.16(1)  \\
$\log g_1$ [cm/s$^2$] & 5.74(5) & 5.77(6) & 5.79(7) \\
$\log g_2$ [cm/s$^2$] & 5.14(5) & 5.14(5) & 5.14(5) \\     
\hline 
\multicolumn{4}{l}{Roche radii: [in units of orbital separation]} \\
\hline  
$r_1$ (pole) & 0.191 & 0.191 & 0.191\\
$r_1$ (point)& 0.193 & 0.193 & 0.193\\
$r_1$ (side) & 0.192 & 0.192 & 0.192\\
$r_1$ (back) & 0.193 & 0.193 & 0.193\\
$r_2$ (pole) & 0.198 & 0.197 & 0.197\\
$r_2$ (point)& 0.213 & 0.215 & 0.216\\
$r_2$ (side) & 0.201 & 0.201 & 0.201\\
$r_2$ (back) & 0.210 & 0.211 & 0.211\\ 
\hline 
\multicolumn{4}{l}{Errors on residuals: } \\
\hline       
$\sigma$($g$') [mag] & 0.03055 & 0.03054 & 0.03057\\
$\sigma$($r$') [mag] & 0.01325 & 0.01321 & 0.01321\\ 
$\sigma$(RV) [km/s] & 8.39 & 8.39 & 8.39 \\ 
\hline       
\end{tabular}
\end{table}

\begin{figure}
\centering
\includegraphics[angle=-90, width=\hsize]{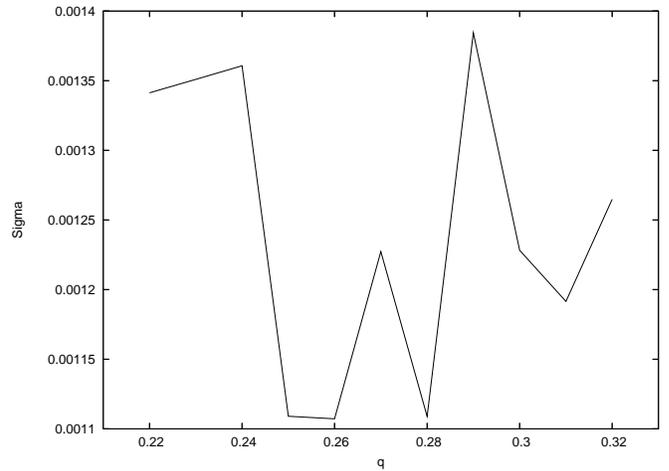}
\caption{ Mass ratio $q$ versus sigma, for the range of the possible $q$
values. Sigma is the sum of the squares of the sigmas in the two considered 
filters
($\sigma$($g$') and $\sigma$($r$')).}
\label{qPOT2_chi2_Gplot}
\end{figure}

As a quantitative measure of the goodness-of-fit we use the 1 $\sigma$ deviation
for each data set ($g$', $r$' and RV) from the simultaneously calculated
synthetic curves. The bigger 1 $\sigma$ deviation in $g$' is due to the higher
amplitudes of the oscillations in this colour.  While it is impossible to see
the depth of the local minima found by the DC method, and therefore search for
the global minimum of the parameter hyperspace, we tested the stability of the
convergent solutions found by parameter kicking \citep{Prsa2005}.  Once
convergence was reached, we manually kicked the parameters and the minimization
was restarted from the displaced points. In this way we found three groups of
solutions of equal goodness-of-fit. Table \ref{tbl:system_param} gives the three
best fit orbital solutions. It is not possible to decide which solution is the
correct one based on the numerical considerations as the synthetic curves are
fitting the data equally well for all three models.  The errors given in the
table are the formal errors of the fit which are likely smaller than the true
errors due to the above mentioned correlation between the parameters.  The
synthetic lightcurve fits to the observed data points are presented in
Fig.~\ref{Fig_rv}, Fig.~\ref{Fig:Gfit} and Fig.~\ref{Fig:Rfit} (solid line)
together with their residuals.  The synthetic $g$' and $r$' lightcurves and the
RV curve are plotted for only one solution (Model II) since the deviations
between the three solutions cannot be resolved at the scale of the figure.

%______________________________________________ 
  
\begin{figure}
\centering
\includegraphics[width=\hsize]{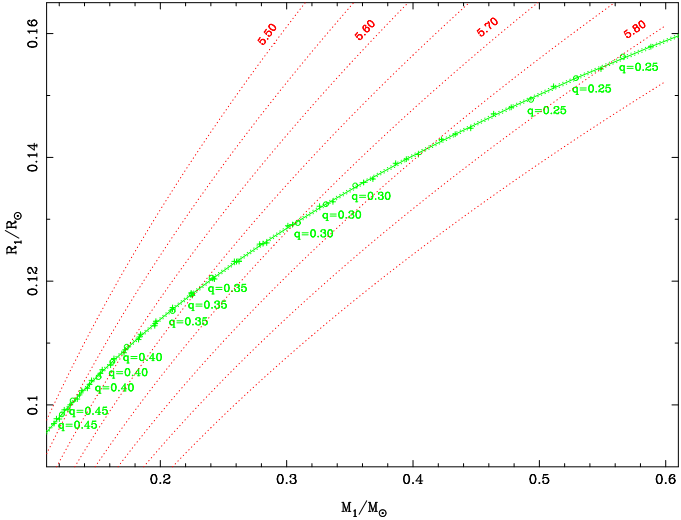}
\caption{
Mass--radius diagram for \nyvir\ showing the regions permitted by the orbit 
solution (continuous line) and by the different surface gravities (dotted
lines). The $q$ values are also noted on the orbit solution.
The small changes from the $3\sigma$ error on $K_1$ do not
shift the curve representing the orbital solution.
}
\label{Fig_massradius}
\end{figure}

\subsection{Discussion}\label{sect:discussion_PG1336}

The uniqueness of a given solution is jeopardized by the parameter
correlations.  In particular, there is a strong correlation between the mass
ratio $q$ and the potential of the secondary star $\Omega_2$. Therefore, there
is a $q$ degeneracy in all the orbital solutions.  For a given range of
potentials defined by the Lagrangian point, a family of solutions with
corresponding mass ratios is found. The solutions found in
Table~\ref{tbl:system_param} represent the local minima shown in
Fig.~\ref{qPOT2_chi2_Gplot}.

The relative radii and the orbital inclination are tightly constrained by the
depth and the width of the eclipses, and the results in all three models are
nearly identical. There is only a slight distortion of the secondary: $r_2$
(pole)/$r_2$ (point) is 0.93, 0.92, 0.91 respectively for each model.  While the
previous searches for the best orbital solutions \cite[][and references
therein]{D.Kilkenny1998MNRAS, Drechsel2001A&A} tend to resort to non--physical
albedos (greater that 1 in some cases) and limb darkening coefficients of the
secondary, we find that the biggest problem is in the temperature of the
secondary which is heated by the hot subdwarf. The weakest point of all
modelling procedures lies in an inadequate treatment of the temperature of the
secondary star. The temperature distribution over the surface of the secondary
has to be incorporated in the atmosphere models used by {\tt PHOEBE} in order to
get more realistic solutions. This is far beyond the scope of our current paper.

The surface gravity derived from the orbital solutions, although in
agreement with the value previously estimated by
\citet[][$\log g  =  5.7 \pm 0.1\,{\rm dex}$]{D.Kilkenny1998MNRAS} 
is higher than the spectroscopic gravity estimate. Therefore, we have 
explored the full range of mass--radius ranges for the primary allowed by 
the orbital solution and the spectroscopic gravity (Fig.~\ref{Fig_massradius}).
The parameters used to generate this orbital solution mass--radius
relationship are only the $P$, $i$, $K_1$ and the radius of the primary
in terms $a$, none of which are affected by the $q$ degeneracy.
Thus, if we had a sufficiently accurate spectroscopic determination of
\logg, we could use the relationships in Fig.~\ref{Fig_massradius}
to determine one unique $M_1$. Unfortunately, our spectroscopic $\log g$ of 5.6
is clearly much lower than what can realistically be accepted since it
gives a mass for the primary that is far too low ($M_1$ $<$ 0.2 [M$_\odot$]).

While we cannot discriminate between the three model fits on the basis of their
$\sigma$ values, the evolutionary scenarios for sdB stars disqualify the Model
III solution as the primary mass would be too high for a core He--flash
\citep{Han2002}. Models I and II however, are both possible as they could have
formed through common envelope phase (Hu et al., submitted to A\&A).
  
%__________________________________________________________________

\section{Detection of the Rossiter-McLaughlin effect}

\begin{figure}
\centering
\includegraphics[angle=-90, width=\hsize]{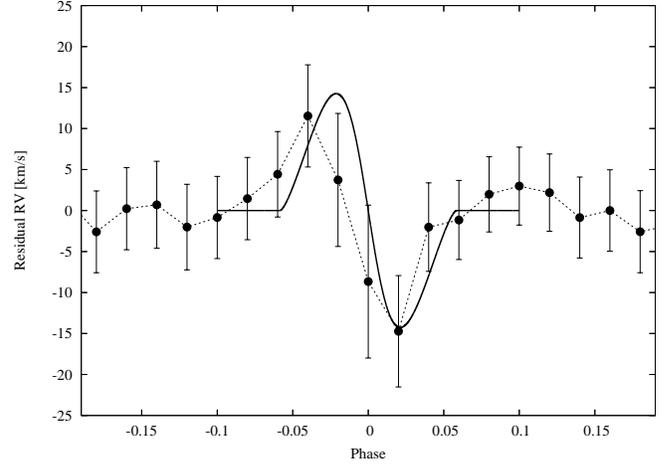}
\caption{The orbit subtracted RV residuals (dots) with their corresponding
errors clearly showing the RM effect. The solid line is the simulation of the RM
effect with the parameters given in the text.}
\label{Fig_RM}
\end{figure}

%__________________________________________________________________

In Fig.\,\ref{Fig_rv}, an apparent up-and-down (redshift-blueshift) shift occurs
at phase zero in the RV curve. This effect at the eclipse is known as the
Rossiter-McLaughlin (RM) effect \citep{Rossiter1924ApJ, McLaughlin1924}. It is
due to the selective blocking of the light of the rotating star during an
eclipse.  When the secondary star covers the blueshifted (redshifted) half of
the stellar disk, the integrated light of the primary appears slightly
redshifted (blueshifted).  Because of this selective blocking of the stellar surface 
during the eclipse,  a skewed line profile is created. This change in line profile shape results in a
shift in RV, which in turn results in the redshift-blueshift distortion seen
during the eclipse (see Fig.~\ref{Fig_rv}). The RM effect has been seen in
other eclipsing hot subdwarf binaries (e.g. AA\,Dor: \cite{Rauch2003}) and can be used to
investigate the rotational properties of the component stars. It was recently
used in extrasolar planetary transits \citep{Queloz2000, Ohta20005,
Giminez2006,Gaudi2007} to discriminate between different migration theories.
The amplitude of the effect mainly depends on the projected rotation velocity of
the star, the ratio of stellar radii, the orbital inclination, and the limb
darkening.

To analyze this effect we have subtracted the orbital solution (solid curve in
Fig.~\ref{Fig_rv}) from the RV measurements.  The orbit-subtracted RV residuals,
phase binned in 50 bins, are plotted in Fig.~\ref{Fig_RM}. The RM effect is
clearly seen in these residuals. We used the analytical description of this
effect given in \cite{Giminez2006} to simulate the RM effect for this system. We have
assumed that the rotational axis of the primary star is
co--aligned with the perpendicular to the orbital plane. The result of this
simulation is plotted as a solid line in Fig.~\ref{Fig_RM}. The equatorial
rotational velocity of the star was set to 75.2\, km\,s$^{-1}$ and the ratio of
the stellar radii $r_2$/$r_1$ , the inclination of the orbit $i$ and the radius of the primary
relative to the size of the orbit $r_1$ were taken from our orbital solution (see
Table ~\ref{tbl:system_param}). The synthetic curve fits the observed RM
amplitude rather well. The uncertainties on the residual RV curve are too large
to fine-tune the orbital parameters. We can only establish that the observed RM
effect is compatible with the orbital solutions given in
Table\,\ref{tbl:system_param} and represents an independent confirmation of the light
curve solution.

The apparent asymmetry seen in Fig.~\ref{Fig_RM} is, however, not well
explained.  Such an asymmetry is expected to occur if the projected orbital and
rotational axes are not aligned. This is highly unlikely for the narrow orbit of
\nyvir. Nevertheless, we simulated the RM effect allowing different angles of
the rotation axes and the orbital axes. We indeed could not achieve satisfactory
results, because, when the zero offset was fitted well, the amplitudes were
highly asymmetrical and vice versa. The asymmetry is more likely caused by the
pulsations seen during the primary eclipse, which also give rise in line profile 
shape variations. The equations
describing the RM effect assume that the components are spherical, i.e.\ they do
not take into account any deviation from spherical symmetry such as the one
produced by the pulsations.  We will investigate this further in
our follow-up paper dedicated to the analysis of the primary's pulsations.
%----------------------------------------------------

\section{Conclusions and Future work}\label{sect:conclusions_PG1336}

In this work, we presented a thorough observational analysis of the orbital
behavior of the pulsating eclipsing binary \nyvir. Our goal was to avoid
using a canonical mass of 0.5\,M$_\odot$ for the subdwarf in any interpretation
of the luminosity variations of the star, as has been done so far in the literature. Instead,
we attempted an unbiased derivation of the system and stellar parameters, in
particular for the masses of the components. Our analysis resulted in three
equally probable sets of orbital and physical parameters of the system. Our
model III solution is incompatible with the binary having gone through a core
He--flash and a common-envelope phase described by the $\alpha$-formalism since
that can only lead to \nyvir\ like binaries with primary masses up to
0.48\,M$_\odot$ (Hu et al., submitted to A\&A).  This leaves us with two
solutions, one with a primary mass of 0.466$\pm$0.006\,M$_\odot$ and another
with 0.389$\pm$0.005\,M$_\odot$, with secondary masses of
0.122$\pm$0.001\,M$_\odot$ and 0.110$\pm$0.001\,M$_\odot$ respectively.  We thus
conclude that our solutions with $M_1=0.466\pm$0.006\,M$_\odot$ and
$M_1=0.389\pm$0.005\,M$_\odot$ are the only plausible ones, except when the
common-envelope phase would be better described by the $\gamma$-formalism
\citep{Nelemans2000,G.Nelemans2005MNRAS}. In this case all three solutions are
acceptable, as this formalism allows non-degenerate helium ignition with a
broader primary mass range (0.3-1.1\,M$_\odot$). 

Furthermore, we have detected the RM effect in the radial velocity curve of
\nyvir. The simulated amplitude of the RM effect is in the accordance with the
RM amplitude seen in the RV residuals, which is an independent confirmation of
the results obtained from our orbital solution.

While deriving the orbital solution for \nyvir, we hit upon the limitation of
current binary analysis codes, which also prevented us to pinpoint the effective
temperature of the secondary.  None of the analysis methods available in the
literature treat the atmosphere of such a close binary, in which one component
is so hot that it induces a temperature gradient across the surface of the
other, in an appropriate way. Indeed, all codes make use of stellar atmosphere
models which assume one fixed effective temperature at the surface of each of
the component stars. As such, any derived quantities, such as limb darkening
coefficients and albedos, cannot be but a very crude approximation of reality
whenever one component is seriously heated by the other one. In the case of
close binaries like \nyvir, i.e.~with a hot primary and a cold secondary, the
temperature of the latter changes so drastically from the illuminated side to
the backside, that specific atmosphere models representing such a situation
should be computed and used while deriving the orbital parameters. This is an
entire project by itself and surely beyond the scope of our current work. We
hope that our results will give rise to future developments of atmosphere models
with temperatures varying across the surface of the cool component in close
binaries. The case of \nyvir, and our data of the star, are ideally suited to
test such new future models.

In a follow-up paper of this work, we plan to analyse the oscillatory signal in
our multicolour photometry and high-resolution spectroscopy, after the orbit
subtraction presented here. This will be done by computing a cross-correlation
function of each spectrum and investigating the signature of the modes in
it. Cross-correlation functions have already been used to study the character of
oscillations modes before, see e.g \cite{20CVdeltaScuti1996} for the
$\delta\,$Scuti star 20\,CVn and \cite{Hekker2006} for solar-like oscillations
in red giants.  This is done by computing line diagnostics, such as moments, and
the amplitude and phase across the profile, and comparing these to predictions
based on the theory of non-radial oscillations. In principle, this allows us to
identify the spherical wavenumbers $(\ell,m)$ of the strongest modes.  The use
of these established mode identification techniques \cite[see e.g.][for the
latest versions]{BriquetandAerts2003,Zima2006I} on high-resolution
cross-correlation profiles of pulsating sdB stars has so far not yet been
done. The nature of our data and of our target star requires a simulation study
to test the effects of smearing out the oscillations over the cycle and of the
limited time base. Also, we must treat the data during and outside the eclipses
separately in order to assess the effectiveness of the techniques in the
specific case of \nyvir. Such a study is currently being performed. The ultimate
goal of it is to identify the highest-amplitude modes and discriminate among the
plausible seismic models of the star. This will then eventually lead us to
derive a seismic mass estimate to be confronted with the observed primary masses
presented here and with the evolutionary masses computed by Hu et al.~(submitted
to A\&A).

\begin{acknowledgements}
      MV thanks Maarten Reyniers for his generous help in the UVES data
      reduction procedure.  MV acknowledges a PhD scholarship from the Research
      Council of Leuven University.  HH acknowledges a PhD scholarship through
      the ``Convenant Katholieke Universiteit Leuven, Belgium -- Radboud
      Universiteit Nijmegen, the Netherlands''. MV, CA, R\O, and HH are
      supported by the Research Council of Leuven University, through grant
      GOA/2003/04. ULTRACAM is supported by PPARC grants PPA/G/S/2003/00058 and
      PP/D002370/1. We are thankfull to Joshua Winn and Scott Gaudi for enlightening discussions
      on the RM effect and to Alvaro Gim{\'e}nez for kindly providing us with
      his subroutines. We thank Prof.\ Uli Heber for kindly providing the LTE
      spectral grids.
\end{acknowledgements}

\bibliographystyle{aa} 
\bibliography{references} 

\begin{thebibliography}{52}
\expandafter\ifx\csname natexlab\endcsname\relax\def\natexlab#1{#1}\fi

\bibitem[{{Aerts} \& {Eyer}(2000)}]{Aerts2000}
{Aerts}, C. \& {Eyer}, L. 2000, in ASP Conf. Ser. 210: Delta Scuti and Related
  Stars, ed. M.~{Breger} \& M.~{Montgomery}, 113

\bibitem[{{Al-Naimiy}(1978)}]{Al-Naimiy1978}
{Al-Naimiy}, H.~M. 1978, \apss, 53, 181

\bibitem[{{Behara} \& {Jeffery}(2006)}]{Behara2006}
{Behara}, N.~T. \& {Jeffery}, C.~S. 2006, \aap, 451, 643

\bibitem[{{Brassard} {et~al.}(2001){Brassard}, {Fontaine}, {Bill{\`e}res},
  {Charpinet}, {Liebert}, \& {Saffer}}]{Brassard2001ApJ}
{Brassard}, P., {Fontaine}, G., {Bill{\`e}res}, M., {et~al.} 2001, \apj, 563,
  1013

\bibitem[{{Briquet} \& {Aerts}(2003)}]{BriquetandAerts2003}
{Briquet}, M. \& {Aerts}, C. 2003, \aap, 398, 687

\bibitem[{{Charpinet} {et~al.}(1996){Charpinet}, {Fontaine}, {Brassard}, \&
  {Dorman}}]{Charpinet1996}
{Charpinet}, S., {Fontaine}, G., {Brassard}, P., \& {Dorman}, B. 1996, \apjl,
  471, L103+

\bibitem[{{Charpinet} {et~al.}(2000){Charpinet}, {Fontaine}, {Brassard}, \&
  {Dorman}}]{Charpinet2000}
{Charpinet}, S., {Fontaine}, G., {Brassard}, P., \& {Dorman}, B. 2000, \apjs,
  131, 223

\bibitem[{{Charpinet} {et~al.}(2005){Charpinet}, {Fontaine}, {Brassard},
  {Green}, \& {Chayer}}]{Charpinet2005A&A}
{Charpinet}, S., {Fontaine}, G., {Brassard}, P., {Green}, E.~M., \& {Chayer},
  P. 2005, \aap, 437, 575

\bibitem[{{Dhillon} \& {Marsh}(2001)}]{Dhillon&Marsh2001}
{Dhillon}, V. \& {Marsh}, T. 2001, New Astronomy Review, 45, 91

\bibitem[{{Dorman} {et~al.}(1993){Dorman}, {Rood}, \&
  {O'Connell}}]{Dorman1993ApJ}
{Dorman}, B., {Rood}, R.~T., \& {O'Connell}, R.~W. 1993, \apj, 419, 596

\bibitem[{{Drechsel} {et~al.}(2001){Drechsel}, {Heber}, {Napiwotzki},
  {{\O}stensen}, {Solheim}, {Johannessen}, {Schuh}, {Deetjen}, \&
  {Zola}}]{Drechsel2001A&A}
{Drechsel}, H., {Heber}, U., {Napiwotzki}, R., {et~al.} 2001, \aap, 379, 893

\bibitem[{{Dreizler} {et~al.}(2000){Dreizler}, {Koester}, \&
  {Heber}}]{Dreizler2000BaltA}
{Dreizler}, S., {Koester}, D., \& {Heber}, U. 2000, Baltic Astronomy, 9, 113

\bibitem[{{Dupret} {et~al.}(2003){Dupret}, {De Ridder}, {De Cat}, {Aerts},
  {Scuflaire}, {Noels}, \& {Thoul}}]{Dupret2003}
{Dupret}, M.-A., {De Ridder}, J., {De Cat}, P., {et~al.} 2003, \aap, 398, 677

\bibitem[{{Fukugita} {et~al.}(1996){Fukugita}, {Ichikawa}, {Gunn}, {Doi},
  {Shimasaku}, \& {Schneider}}]{Fukugita1996}
{Fukugita}, M., {Ichikawa}, T., {Gunn}, J.~E., {et~al.} 1996, \aj, 111, 1748

\bibitem[{{Fusi-Pecci} \& {Renzini}(1976)}]{Fusi-Pecci1976A&A}
{Fusi-Pecci}, F. \& {Renzini}, A. 1976, \aap, 46, 447

\bibitem[{{Gaudi} \& {Winn}(2006)}]{Gaudi2007}
{Gaudi}, B.~S. \& {Winn}, J.~N. 2006, ArXiv Astrophysics e-prints

\bibitem[{{Gim{\'e}nez}(2006)}]{Giminez2006}
{Gim{\'e}nez}, A. 2006, \apj, 650, 408

\bibitem[{{Green} {et~al.}(2003){Green}, {Fontaine}, {Reed}, {Callerame},
  {Seitenzahl}, {White}, {Hyde}, {{\O}stensen}, {Cordes}, {Brassard}, {Falter},
  {Jeffery}, {Dreizler}, {Schuh}, {Giovanni}, {Edelmann}, {Rigby}, \&
  {Bronowska}}]{Green2003}
{Green}, E.~M., {Fontaine}, G., {Reed}, M.~D., {et~al.} 2003, \apjl, 583, L31

\bibitem[{{Green} {et~al.}(1986){Green}, {Schmidt}, \&
  {Liebert}}]{GreenSchmidtandLiebert86}
{Green}, R.~F., {Schmidt}, M., \& {Liebert}, J. 1986, \apjs, 61, 305

\bibitem[{{Han} {et~al.}(2003){Han}, {Podsiadlowski}, {Maxted}, \&
  {Marsh}}]{Han2003}
{Han}, Z., {Podsiadlowski}, P., {Maxted}, P.~F.~L., \& {Marsh}, T.~R. 2003,
  \mnras, 341, 669

\bibitem[{{Han} {et~al.}(2002){Han}, {Podsiadlowski}, {Maxted}, {Marsh}, \&
  {Ivanova}}]{Han2002}
{Han}, Z., {Podsiadlowski}, P., {Maxted}, P.~F.~L., {Marsh}, T.~R., \&
  {Ivanova}, N. 2002, \mnras, 336, 449

\bibitem[{{Heber} {et~al.}(2000){Heber}, {Reid}, \& {Werner}}]{Heber2000}
{Heber}, U., {Reid}, I.~N., \& {Werner}, K. 2000, \aap, 363, 198

\bibitem[{{Hekker} {et~al.}(2006){Hekker}, {Aerts}, {de Ridder}, \&
  {Carrier}}]{Hekker2006}
{Hekker}, S., {Aerts}, C., {de Ridder}, J., \& {Carrier}, F. 2006, \aap, 458,
  931

\bibitem[{{Jeffery} {et~al.}(2005){Jeffery}, {Aerts}, {Dhillon}, {Marsh}, \&
  {G{\"a}nsicke}}]{Jeffery2005}
{Jeffery}, C.~S., {Aerts}, C., {Dhillon}, V.~S., {Marsh}, T.~R., \&
  {G{\"a}nsicke}, B.~T. 2005, \mnras, 362, 66

\bibitem[{{Kilkenny} {et~al.}(2000){Kilkenny}, {Keuris}, {Marang}, {Roberts},
  {van Wyk}, \& {Ogloza}}]{D.Kilkenny2000Obs}
{Kilkenny}, D., {Keuris}, S., {Marang}, F., {et~al.} 2000, The Observatory,
  120, 48

\bibitem[{{Kilkenny} {et~al.}(1997){Kilkenny}, {Koen}, {O'Donoghue}, \&
  {Stobie}}]{Kilkenny1997}
{Kilkenny}, D., {Koen}, C., {O'Donoghue}, D., \& {Stobie}, R.~S. 1997, \mnras,
  285, 640

\bibitem[{{Kilkenny} {et~al.}(1998){Kilkenny}, {O'Donoghue}, {Koen},
  {Lynas-Gray}, \& {van Wyk}}]{D.Kilkenny1998MNRAS}
{Kilkenny}, D., {O'Donoghue}, D., {Koen}, C., {Lynas-Gray}, A.~E., \& {van
  Wyk}, F. 1998, \mnras, 296, 329

\bibitem[{{Kilkenny} {et~al.}(2003){Kilkenny}, {Reed}, {O'Donoghue}, {Kawaler},
  {Mukadam}, {Kleinman}, {Nitta}, {Metcalfe}, {Provencal}, {Watson},
  {Sullivan}, {Sullivan}, {Shobbrook}, {Jiang}, {Joshi}, {Ashoka}, {Seetha},
  {Leibowitz}, {Ibbetson}, {Mendelson}, {Mei{\v s}tas}, {Kalytis}, {Ali{\v
  s}auskas}, {Martinez}, {van Wyk}, {Stobie}, {Marang}, {Zola}, {Krzesinski},
  {Og{\l}oza}, {Moskalik}, {Silvotti}, {Piccioni}, {Vauclair}, {Dolez},
  {Chevreton}, {Dreizler}, {Schuh}, {Deetjen}, {Solheim}, {Gonzalez Perez},
  {Ulla}, {{\O}stensen}, {Manteiga}, {Suarez}, {Burleigh}, {Kepler}, {Kanaan},
  \& {Giovannini}}]{D.Kilkenny2003MNRAS}
{Kilkenny}, D., {Reed}, M.~D., {O'Donoghue}, D., {et~al.} 2003, \mnras, 345,
  834

\bibitem[{{Lenz} \& {Breger}(2004)}]{Period04}
{Lenz}, P. \& {Breger}, M. 2004, in IAU Symposium, ed. J.~e.~a. {Zverko},
  786--790

\bibitem[{{Mathias} \& {Aerts}(1996)}]{20CVdeltaScuti1996}
{Mathias}, P. \& {Aerts}, C. 1996, \aap, 312, 905

\bibitem[{{Maxted} {et~al.}(2001){Maxted}, {Heber}, {Marsh}, \&
  {North}}]{Maxted_2001}
{Maxted}, P.~f.~L., {Heber}, U., {Marsh}, T.~R., \& {North}, R.~C. 2001,
  \mnras, 326, 1391

\bibitem[{{McLaughlin}(1924)}]{McLaughlin1924}
{McLaughlin}, D.~B. 1924, \apj, 60, 22

\bibitem[{{Menzies} \& {Marang}(1986)}]{Menzies1986}
{Menzies}, J.~W. \& {Marang}, F. 1986, in IAU Symp. 118: Instrumentation and
  Research Programmes for Small Telescopes, ed. J.~B. {Hearnshaw} \& P.~L.
  {Cottrell}, 305

\bibitem[{{Morales-Rueda} {et~al.}(2006){Morales-Rueda}, {Maxted}, {Marsh},
  {Kilkenny}, \& {O'Donoghue}}]{Morales-Rueda2006ECSurvey}
{Morales-Rueda}, L., {Maxted}, P.~F.~L., {Marsh}, T.~R., {Kilkenny}, D., \&
  {O'Donoghue}, D. 2006, Baltic Astronomy, 15, 187

\bibitem[{{Morales-Rueda} {et~al.}(2003){Morales-Rueda}, {Maxted}, {Marsh},
  {North}, \& {Heber}}]{M-Rueda2003MNRAS}
{Morales-Rueda}, L., {Maxted}, P.~F.~L., {Marsh}, T.~R., {North}, R.~C., \&
  {Heber}, U. 2003, \mnras, 338, 752

\bibitem[{{Mukai}(1990)}]{Mukai1990}
{Mukai}, K. 1990, \pasp, 102, 183

\bibitem[{{Napiwotzki} {et~al.}(2004){Napiwotzki}, {Karl}, {Lisker}, {Heber},
  {Christlieb}, {Reimers}, {Nelemans}, \& {Homeier}}]{Napiwotzki2004SPY}
{Napiwotzki}, R., {Karl}, C.~A., {Lisker}, T., {et~al.} 2004, \apss, 291, 321

\bibitem[{{Nelemans} \& {Tout}(2005)}]{G.Nelemans2005MNRAS}
{Nelemans}, G. \& {Tout}, C.~A. 2005, \mnras, 356, 753

\bibitem[{{Nelemans} {et~al.}(2000){Nelemans}, {Verbunt}, {Yungelson}, \&
  {Portegies Zwart}}]{Nelemans2000}
{Nelemans}, G., {Verbunt}, F., {Yungelson}, L.~R., \& {Portegies Zwart}, S.~F.
  2000, \aap, 360, 1011

\bibitem[{{Ohta} {et~al.}(2005){Ohta}, {Taruya}, \& {Suto}}]{Ohta20005}
{Ohta}, Y., {Taruya}, A., \& {Suto}, Y. 2005, \apj, 622, 1118

\bibitem[{{{\O}stensen} {et~al.}(2007){{\O}stensen}, {Oreiro}, {Drechsel},
  {Heber}, \& {Pigulski}}]{Roy2007}
{{\O}stensen}, R., {Oreiro}, R., {Drechsel}, H., {Heber}, U., \& {Pigulski}, A.
  2007, in ASP Conf.~Ser.: 15th European Workshop on White Dwarfs, ed.
  R.~{Napiwotzki}, in press.

\bibitem[{{Pr{\v s}a} \& {Zwitter}(2005)}]{Prsa2005}
{Pr{\v s}a}, A. \& {Zwitter}, T. 2005, \apj, 628, 426

\bibitem[{{Queloz} {et~al.}(2000){Queloz}, {Eggenberger}, {Mayor}, {Perrier},
  {Beuzit}, {Naef}, {Sivan}, \& {Udry}}]{Queloz2000}
{Queloz}, D., {Eggenberger}, A., {Mayor}, M., {et~al.} 2000, \aap, 359, L13

\bibitem[{{Randall} {et~al.}(2005){Randall}, {Fontaine}, {Brassard}, \&
  {Bergeron}}]{Randall2005ApJS}
{Randall}, S.~K., {Fontaine}, G., {Brassard}, P., \& {Bergeron}, P. 2005,
  \apjs, 161, 456

\bibitem[{{Rauch} \& {Werner}(2003)}]{Rauch2003}
{Rauch}, T. \& {Werner}, K. 2003, \aap, 400, 271

\bibitem[{{Rossiter}(1924)}]{Rossiter1924ApJ}
{Rossiter}, R.~A. 1924, \apj, 60, 15

\bibitem[{{Wade} \& {Rucinski}(1985)}]{Wade1985}
{Wade}, R.~A. \& {Rucinski}, S.~M. 1985, \aaps, 60, 471

\bibitem[{{Wilson} \& {Devinney}(1971)}]{WilsonandDevinney1971}
{Wilson}, R.~E. \& {Devinney}, E.~J. 1971, \apj, 166, 605

\bibitem[{{Wood} {et~al.}(1993){Wood}, {Zhang}, \& {Robinson}}]{Wood1993}
{Wood}, J.~H., {Zhang}, E.-H., \& {Robinson}, E.~L. 1993, \mnras, 261, 103

\bibitem[{{Woolf} {et~al.}(2003){Woolf}, {Jeffery}, \& {Pollacco}}]{Woolf2003}
{Woolf}, V.~M., {Jeffery}, C.~S., \& {Pollacco}, D. 2003, in NATO ASIB Proc.
  105: White Dwarfs, ed. D.~e.~a. {de Martino}, 95

\bibitem[{{Zima}(2006)}]{Zima2006I}
{Zima}, W. 2006, \aap, 455, 227

\bibitem[{{Zola}(2000)}]{S.Zola2000BaltA}
{Zola}, S. 2000, Baltic Astronomy, 9, 197

\end{thebibliography}
\end{document}